





\message{--Loading JNL macros, please be patient--}
\font\twelverm=cmr12    \font\twelvei=cmmi12
\font\twelvesy=cmsy10 scaled 1200   \font\twelveex=cmex10 scaled 1200
\font\twelvebf=cmbx12   \font\twelvesl=cmsl12
\font\twelvett=cmtt12   \font\twelveit=cmti12

\skewchar\twelvei='177   \skewchar\twelvesy='60


\def\twelvepoint{\normalbaselineskip=12.4pt
  \abovedisplayskip 12.4pt plus 3pt minus 9pt
  \belowdisplayskip 12.4pt plus 3pt minus 9pt
  \abovedisplayshortskip 0pt plus 3pt
  \belowdisplayshortskip 7.2pt plus 3pt minus 4pt
  \smallskipamount=3.6pt plus1.2pt minus1.2pt
  \medskipamount=7.2pt plus2.4pt minus2.4pt
  \bigskipamount=14.4pt plus4.8pt minus4.8pt
  \def\rm{\fam0\twelverm}          \def\it{\fam\itfam\twelveit}%
  \def\sl{\fam\slfam\twelvesl}     \def\bf{\fam\bffam\twelvebf}%
  \def\mit{\fam 1}                 \def\cal{\fam 2}%
  \def\tt{\twelvett}
  \textfont0=\twelverm   \scriptfont0=\tenrm   \scriptscriptfont0=\sevenrm
  \textfont1=\twelvei    \scriptfont1=\teni    \scriptscriptfont1=\seveni
  \textfont2=\twelvesy   \scriptfont2=\tensy   \scriptscriptfont2=\sevensy
  \textfont3=\twelveex   \scriptfont3=\twelveex  \scriptscriptfont3=\twelveex
  \textfont\itfam=\twelveit
  \textfont\slfam=\twelvesl
  \textfont\bffam=\twelvebf \scriptfont\bffam=\tenbf
  \scriptscriptfont\bffam=\sevenbf
  \normalbaselines\rm}

\font\twelverms=cmr10 scaled 2073
\font\twelvebfs=cmbx10 scaled 2073
\font\twelveits=cmti10 scaled 2073
\font\twelvesls=cmsl10 scaled 2073

\def\biggertype{\let\rm=\twelverms
	\let\bf=\twelvebfs
	\let\it=\twelveits
	\let\sl=\twelvesls
	\rm}

\font\twelverms=cmr10 scaled 1728
\font\twelvebfs=cmbx10 scaled 1728
\font\twelveits=cmti10 scaled 1728
\font\twelvesls=cmsl10 scaled 1728

\def\biggertype{\let\rm=\twelverms
	\let\bf=\twelvebfs
	\let\it=\twelveits
	\let\sl=\twelvesls
	\baselineskip=17pt minus 1pt
	\rm}



                 \font\ninei=cmmi9
\font\ninesy=cmsy9

\skewchar\ninei='177   \skewchar\ninesy='60



\def\beginlinemode{\endmode
  \begingroup\parskip=0pt \obeylines\def\\{\par}\def\endmode{\par\endgroup}}
\def\beginparmode{\endmode
  \begingroup \def\endmode{\par\endgroup}}
\let\endmode=\par
{\obeylines\gdef\
{}}
\def\singlespace{\baselineskip=\normalbaselineskip}

\def\oneandahalfspace{\baselineskip=\normalbaselineskip
  \multiply\baselineskip by 3 \divide\baselineskip by 2}
\def\doublespace{\baselineskip=\normalbaselineskip \multiply\baselineskip by 2}

\newcount\firstpageno
\firstpageno=2
\footline={\ifnum\pageno<\firstpageno{\hfil}
  \else{\hfil\twelverm\folio\hfil}\fi}
\let\rawfootnote=\footnote		
\def\footnote#1#2{{\rm\singlespace\parindent=0pt\rawfootnote{#1}{#2}}}
\def\raggedcenter{\leftskip=4em plus 12em \rightskip=\leftskip
  \parindent=0pt \parfillskip=0pt \spaceskip=.3333em \xspaceskip=.5em
  \pretolerance=9999 \tolerance=9999
  \hyphenpenalty=9999 \exhyphenpenalty=9999 }
\def\dateline{\rightline{\ifcase\month\or
  January\or February\or March\or April\or May\or June\or
  July\or August\or September\or October\or November\or December\fi
  \space\number\year}}
\def\received{\vskip 3pt plus 0.2fill
 \centerline{\sl (Received\space\ifcase\month\or
  January\or February\or March\or April\or May\or June\or
  July\or August\or September\or October\or November\or December\fi
  \qquad, \number\year)}}


\hsize=6.5truein
\hoffset=0truein
\vsize=8.1truein
\voffset=0truein
\parskip=\medskipamount
\twelvepoint		
\doublespace		
\overfullrule=0pt	



\def\title			
  {\null\vskip 3pt plus 0.2fill
   \beginlinemode \doublespace \raggedcenter \bf}

\def\author			
  {\vskip 3pt plus 0.2fill \beginlinemode
   \singlespace \raggedcenter}

\def\affil			
  {\vskip 3pt plus 0.1fill \beginlinemode
   \oneandahalfspace \raggedcenter \sl}

\def\abstrtitle#1{
   \gdef\actualabstrtitle{#1}\relax
   \xdef\abstrnext{\def\noexpand\obstrtit{\actualabstrtitle}}\relax
   \uppercase\expandafter{\abstrnext}\relax
   \message{ -- Abstract Title set to \obstrtit\space -- }}

\abstrtitle{abstract: }

\def\abstract			
  {\vskip 3pt plus 0.3fill \beginparmode
   \doublespace \narrower }

\def\endtitlepage		
  {\endpage			
   \body}

\def\body			
  {\beginparmode}		

\def\head#1{			
  \filbreak\vskip 0.5truein	
  {\immediate\write16{#1}
   \raggedcenter \expandafter\uppercase{#1}\par}
   \nobreak\vskip 0.25truein\nobreak}

\def\refto#1{$^{#1}$}		

\def\reftitle#1{
   \gdef\actualtitle{#1}\relax
   \xdef\refnext{\def\noexpand\roftit{\actualtitle}}\relax
   \uppercase\expandafter{\refnext}\relax
   \message{ -- Reference Title set to \roftit\space -- }}

\reftitle{References	
  }

\def\references			
  {\head{\roftit}		
commas).
   \beginparmode
   \frenchspacing \parindent=0pt \leftskip=1truecm
   \parskip=8pt plus 3pt \everypar{\hangindent=\parindent}}

\gdef\refis#1{\indent\hbox to 0pt{\hss#1.~}}	

\gdef\journal#1, #2, #3 1#4#5#6{		
sets
    {\sl #1~}{\bf #2}, #3 (1#4#5#6)}		

\gdef\zeitschrift#1, #2, #3, #4 1#5#6#7{	
sets
    {\sl #1~}{\bf #2}, #3, (#4 1#5#6#7)}	
\gdef\newpaper#1, #2, #3,  1#4#5#6{	        
sets
    {\sl #1~}{\bf #2},  (#3, 1#4#5#6)}	        

\gdef\topublish#1, {                            
    to be published in {\sl #1~}}               

\def\pra{\journal Phys. Rev. A, }

\def\prl{\journal Phys. Rev. Lett., }

\def\jpf:mp{\journal J. Phys. F: Met. Phys., }

\def\jfm{\journal J. Fluid Mech., }

\def\physs{\journal Phys. Script., }

\def\endreferences{\body}

\def\figtitle#1{
   \gdef\actualfigtitle{#1}\relax
   \xdef\fignext{\def\noexpand\fogtit{\actualfigtitle}}\relax
   \uppercase\expandafter{\fignext}\relax
   \message{ -- Figure Caption Title set to \fogtit\space -- }}

\figtitle{		
 }

\fogtit{ }

\def\figurecaptions		
  {\endpage
   \beginparmode
   \head{\fogtit Figure Captions}
   \parskip=24pt plus 3pt \everypar={\hangindent=4em}
}

\def\endfigurecaptions{\body}

\def\endreferences{\body}

\def\figurecaptions		
  {\endpage
   \beginparmode
   \head{Figure Captions}
}

\def\endfigurecaptions{\body}

\def\tablecaptions		
  {\endpage
   \beginparmode
   \head{Table Captions}
}

\def\endpage			
  {\vfill\eject}

\def\endpaper			
  {\endmode\vfill\supereject}

\def\endit
  {\endpaper\end}


\def\ref#1{Ref. $#1$}			
\def\Ref#1{Ref. $#1$}			
\def\refnd#1{$#1$}			
\def\Refnd#1{$#1$}			
\def\refhide#1{$#1$} 			


\def\frac#1#2{{\textstyle{#1 \over #2}}}

\def\sla{\raise.15ex\hbox{$/$}\kern-.57em}
\def\leaderfill{\leaders\hbox to 1em{\hss.\hss}\hfill}
\def\twiddle{\lower.9ex\rlap{$\kern-.1em\scriptstyle\sim$}}
\def\bigtwiddle{\lower1.ex\rlap{$\sim$}}
\def\gtwid{\mathrel{\raise.3ex\hbox{$>$\kern-.75em\lower1ex\hbox{$\sim$}}}}
\def\ltwid{\mathrel{\raise.3ex\hbox{$<$\kern-.75em\lower1ex\hbox{$\sim$}}}}
\def\square{\kern1pt\vbox{\hrule height 1.2pt\hbox{\vrule width 1.2pt\hskip 3pt
   \vbox{\vskip 6pt}\hskip 3pt\vrule width 0.6pt}\hrule height 0.6pt}\kern1pt}

\def\today{\ifcase\month\or January\or February\or March\or April\or May\or
  June\or July\or August\or September\or October\or November\or December\fi
  \space\number\day, \number\year}
\def\heute{\number\day. \ifcase\month\or Januar\or Februar\or M\"arz\or
  April\or Mai\or Juni\or Juli\or August\or September\or Oktober\or
  November\or Dezember\fi \space\number\year}

\catcode`@=11
\newcount\r@fcount \r@fcount=0
\newcount\r@@f@count \r@@f@count=0
\newcount\r@fcurr
\newcount\@@total \@@total=0
\immediate\newwrite\reffile
\newif\ifr@ffile\r@ffilefalse
\def\w@rnwrite#1{\ifr@ffile\immediate\write\reffile{#1}\fi\message{#1}}

\def\writer@f#1>>{}
\def\referencefile{
\r@ffiletrue\immediate\openout\reffile=\jobname.ref%
\def\writer@f##1>>{\ifr@ffile\immediate\write\reffile%
{\noexpand\refis{##1} = \csname r@fnum##1\endcsname = %
\expandafter\expandafter\expandafter\strip@t\expandafter%
\meaning\csname r@ftext\csname r@fnum##1\endcsname\endcsname}\fi}%
\def\strip@t##1>>{}}

\def\citeall#1{\xdef#1##1{#1{\noexpand\cite{##1}}}}

\def\citeallnp#1{\xdef#1##1{#1{\noexpand\citenp{##1}}}}	

\def\citenp#1{{\global\@@total = 0}
\each@rgnp\citer@ngenp{#1}}

\def\cite@#1{{\global\@@total = 0}
\begingroup{\each@rg\citer@nge{#1}}
 \aftergroup\@@@sort\aftergroup\@@printit\endgroup
}
\def\cite@@#1{\each@rg\citer@nge{#1}}

\def\each@rg#1#2{{\let\thecsname=#1\expandafter\first@rg#2,\end,}}
\def\each@rgnp#1#2{{\let\thecsname=#1\expandafter\first@rgnp#2,\end,}}

\def\first@rg#1,{\thecsname{#1}\apply@rg}	
\def\first@rgnp#1,{\thecsname{#1}\apply@rg@}	

\def\apply@rg@#1,{\ifx\end#1\let\next=\relax
\else\thecsname{#1}\let\next=\apply@rg@\fi\next}
\def\apply@rg@@#1,{\ifx\end#1\let\next=\relax
\else,\thecsname{#1}\let\next=\apply@rg@@\fi\next}

\def\citer@nge#1{\citedor@nge#1-\end-}	
\def\citer@ngenp#1{\citedor@ngenp#1-\end-}	
\let\@@citedorange = \citer@ange
\def\citerange#1{\expandafter\@@citedorange#1}
\def\citer@ngeat#1\end-{#1}
\def\citedor@nge#1-#2-{\ifx\end#2\r@featspace#1 
\else\citel@@p{#1}{#2}\citer@ngeat\fi}	

\def\citedor@ngenp#1-#2-{\ifx\end#2\r@featspacenp#1 
\else\citel@@p@{#1}{#2}\citer@ngeat\fi}	

\def\citel@@p@#1#2{\ifnum#1>#2{\errmessage{Reference range #1-#2\space is bad.}
\errhelp{If you cite a series of references by the notation M-N, then M and
N must be integers, and N must be greater than or equal to M.}}\else%
{\count0=#1\count1=#2\advance\count1 by1\relax\relax
\expandafter\r@fcite\the\count0,%
\loop\advance\count0 by1\relax
\ifnum\count0<\count1
\expandafter\r@fcite\the\count0,\repeat}\fi}

\def\citel@@p@@#1#2{\ifnum#1>#2{\errmessage{Reference range #1-#2\space is
bad.}
\errhelp{If you cite a series of references by the notation M-N, then M and
N must be integers, and N must be greater than or equal to M.}}\else%
{\count0=#1\count1=#2\advance\count1 by1\relax\expandafter\r@fcite\the\count0,%
\loop\advance\count0 by1\relax
\ifnum\count0<\count1,\expandafter\r@fcite\the\count0,
\repeat}\fi}

\def\r@featspace#1#2 {\r@fcite#1#2,}	

\def\r@featspacenp#1#2 {\r@fcite@#1#2,}	

\def\r@fcite@#1,{\@@inctotal\ifuncit@d{#1}
\expandafter\gdef\csname r@ftext\number\r@fcount\endcsname%
{\message{Reference #1 to be supplied.}\writer@f#1>>#1 to be supplied.\par}\fi%
\expandafter\xdef\csname @@refnum\number\@@total\endcsname{
\number\csname r@fnum#1\endcsname}
}

\def\r@fcite@@#1,{\ifuncit@d{#1}
\expandafter\gdef\csname r@ftext\number\r@fcount\endcsname%
{\message{Reference #1 to be supplied.}\writer@f#1>>#1 to be supplied.\par}\fi%
\csname r@fnum#1\endcsname
}



\def\@@inctotal{
{\global\advance\@@total1\relax}
}

\newcount\@@endn \@@endn=0
\newcount\@@i \@@i=0
\newcount\@@endnin \@@endnin=0
\newcount\@@j \@@j=0
\newcount\@@ct \@@ct=0
\newcount\@@ii \@@ii=0
\newcount\@@temp \@@temp=0
\newcount\@@iii \@@iii=0

\def\ifuncit@d#1{\expandafter\ifx\csname r@fnum#1\endcsname\relax%
\global\advance\r@fcount1\relax%
\expandafter\xdef\csname r@fnum#1\endcsname{\number\r@fcount}}

\def\@@@sort{
{\global\@@endn=\@@total}
{\advance\@@endn-1}
{\global\@@i = 0}
{\loop\ifnum\@@i<\@@endn
\advance\@@i1
\@@@sortin\@@i
\repeat}
}

\def\@@@sortin#1{
{\global\@@endnin = \@@total}{\global\@@j = #1}
{\loop\ifnum\@@j<\@@endnin
\advance\@@j1\relax\@@sortina#1,\@@j,
\repeat}}

\newcount\@@tempa \@@tempa=0
\newcount\@@tempb \@@tempb=0
\newcount\@@tempc \@@tempc=0
\newcount\@@tempd \@@tempd=0
\def\@@sortina#1,#2,{
\@@tempc=#1
\@@tempd=#2
\@@tempa=\csname @@refnum\number\@@tempc\endcsname
\@@tempb=\csname @@refnum\number\@@tempd\endcsname
\relax
\ifnum\@@tempa>\@@tempb
\@@temp=\@@tempa
\@@tempa=15
{\expandafter\xdef\csname @@refnum\number\@@tempc\endcsname{\number\@@tempb}}
{\expandafter\xdef\csname @@refnum\number\@@tempd\endcsname{\number\@@temp}}
\fi}



\def\@@printit{
{\global\@@ct = 0}
{\global\@@i = 0}
{\global\@@endn = \@@total}
{\loop\ifnum\@@i<\@@endn
\advance\@@i1\relax
\@@decideit\@@i,
\repeat}
\@@printitc
\message{)}
}

\def\@@decideit#1,{
\@@temp=1
\ifnum\number#1=\@@temp
{\csname @@refnum1\endcsname}
\message{  Ref(\csname @@refnum1\endcsname}
\else
\@@printita#1,
\fi
}

\newcount\@@ii \@@ii=0
\def\@@printita#1,{
\@@ii=#1
\@@tempc=#1
\advance\@@ii-1\relax
\@@temp=\csname @@refnum\number\@@ii\endcsname
\@@tempb=\csname @@refnum\number#1\endcsname
\advance\@@temp1\relax
\ifnum\@@tempb=\@@temp
\global\advance\@@ct1\relax
\else
\@@printitb#1,
\fi
}

\def\@@printitb#1,{
\@@iii=#1
\advance\@@iii-1\relax
\ifnum\@@ct=0
,\expandafter\number\csname @@refnum\number#1\endcsname
\message{,\expandafter\number\csname @@refnum\number#1\endcsname}
\else
\@@printitbb\@@iii,#1,
\fi
\global\@@ct=0
}

\def\@@printitbb#1,#2,{
\ifnum\@@ct>1
-\expandafter\number\csname @@refnum\number#1\endcsname,
\expandafter\number\csname @@refnum\number#2\endcsname
\message{-\expandafter\number\csname @@refnum\number#1\endcsname,
\expandafter\number\csname @@refnum\number#2\endcsname}
\else
,\expandafter\number\csname @@refnum\number#1\endcsname,
\expandafter\number\csname @@refnum\number#2\endcsname
\message{,\expandafter\number\csname @@refnum\number#1\endcsname,
\expandafter\number\csname @@refnum\number#2\endcsname}
\fi
}

\def\@@printitc{
\ifnum\@@ct=1
,\expandafter\number\csname @@refnum\number\@@total\endcsname
\message{,\expandafter\number\csname @@refnum\number\@@total\endcsname}
\else
\@@printitd
\fi
}

\def\@@printitd{
\ifnum\@@ct>1
-\expandafter\number\csname @@refnum\number\@@total\endcsname
\message{-\expandafter\number\csname @@refnum\number\@@total\endcsname}
\fi
}



\let\r@fis=\refis
\def\r@@fis#1#2#3\par{\ifuncit@d{#1}
\w@rnwrite{.......Reference #1=\number\r@fcount\space is not cited up to now. }
\else%
\fi
\expandafter\gdef\csname r@ftext\csname r@fnum#1\endcsname\endcsname
{\writer@f#1>>#2#3\par}}

\def\r@@@fis#1#2#3\par{\ifuncit@d{#1}  
\else%
\expandafter\gdef\csname r@ftext\csname r@fnum#1\endcsname\endcsname
{\writer@f#1>>#2#3\par}
\fi   }

\def\r@ferr{\endreferences\errmessage{I was expecting to see
\noexpand\endreferences before now;  I have inserted it here.}}
\let\r@ferences=\references
\def\r@f@r@nc@s{\r@ferences\def\endmode{\r@ferr\par\endgroup}}

\let\endr@ferences=\endreferences
\def\endr@f@r@nc@s{\r@fcurr=0
 {\loop\ifnum\r@fcurr<\r@@f@count
    \advance\r@fcurr by 1\relax\expandafter\r@fis\expandafter{\number\r@fcurr}%
    \csname r@ftext\number\r@fcurr\endcsname%
  \repeat}\gdef\r@ferr{}\endr@ferences}


\let\r@fend=\endpaper\gdef\endpaper{\ifr@ffile
\immediate\write16{Cross References written on []\jobname.REF.}\fi\r@fend}


\def\shownumber{\message{--------refcount  is \number\r@fcount\space}
                \message{========refcount1 is \number\r@@f@count\space}}


\def\printallreferences{
\global\let\refis=\r@@fis
\global\let\references=\r@f@r@nc@s
\def\endreferences{
{\global\r@@f@count=\r@fcount\relax}
\endr@f@r@nc@s}
\message{--List all references--}
}

\let\refis=\r@@fis
\let\references=\r@f@r@nc@s
\def\endreferences{
\r@@f@count=\r@fcount
\endr@f@r@nc@s}

\def\printcitedreferences{
\global\let\endreferences=\endr@f@r@nc@s
\gdef\references{
\global\r@@f@count=\r@fcount\relax
\r@f@r@nc@s}
\global\let\refis=\r@@@fis
\message{--List only cited references--}
}

\def\expandedcitations{
\global\let\cite=\cite@@
\global\let\r@fcite=\r@fcite@@
\global\let\citel@@p=\citel@@p@@
\global\let\apply@rg=\apply@rg@@
\message{--Expand all citations--}
}

\def\normalcitations{
\global\let\cite=\cite@
\global\let\r@fcite=\r@fcite@
\global\let\citel@@p=\citel@@p@
\global\let\apply@rg=\apply@rg@
\message{--Order and concentrate citations--}
}

\let\cite=\cite@@
\let\r@fcite=\r@fcite@@
\let\citel@@p=\citel@@p@@
\let\apply@rg=\apply@rg@@

\catcode`@=12

\citeall\refto		
\citeall\ref		%
\citeall\Ref		%
\citeall\refnd		%
\citeall\Refnd		%
\citeallnp\refhide	

\def\refstylenp{		
  \gdef\refto##1{ $[##1]$}				
  \citeall\refto
  \gdef\refis##1{\indent\hbox to 0pt{\hss##1)~}}	
  \gdef\journal##1, ##2, ##3, ##4 {			
     {\sl ##1~}{\bf ##2~}(##3) ##4 }}

\def\refstyleprnp{		
  \gdef\refto##1{ $[##1]$}				
  \citeall\refto
  \gdef\refis##1{\indent\hbox to 0pt{\hss##1)~}}	
  \gdef\journal##1, ##2, ##3, 1##4##5##6{		
    {\sl ##1~}{\bf ##2~}(1##4##5##6) ##3}}

\printcitedreferences
\normalcitations
\referencefile
\title
            Spiral Defect Chaos in Large Aspect
                  Ratio Rayleigh-B\'enard Convection

\author           Stephen W. Morris,$^*$ Eberhard Bodenschatz,$^{\dagger}$
                  David S. Cannell, and Guenter Ahlers

\affil            Department of Physics and
                  Center for Nonlinear Science
                  University of California
                  Santa Barbara, CA 93106-9530

\abstract

We report experiments on convection patterns in a cylindrical cell with
 a large aspect ratio.  The fluid had a Prandtl number ${\sigma  \approx 1}$.
 We observed a chaotic pattern consisting of many rotating spirals and other
defects in the parameter range where theory predicts that steady straight rolls
should be stable.  The correlation length of the pattern decreased rapidly with
 increasing control parameter so that the size of a correlated area became
much smaller than the area of the cell.  This suggests that the chaotic
behavior is intrinsic to large aspect ratio geometries.

{\it Submitted to Phys. Rev. Lett.  May 12 1993.}

\vfill
\line{\quad\quad PACS No. 47.20.Bp\hfil}
\endtitlepage

\doublespace
Rayleigh-B\'enard convection, the instability of a horizontal
fluid layer heated from below,
has served as a paradigm for the study of nonlinear pattern
 formation in systems under
nonequilibrium conditions.\refto{CH93}  One important reason
for this is the extensive nonlinear stability analysis that has been carried
out
by Busse and Clever,\refto{CB74,BC79a} which provides an unusually
detailed picture of the secondary instabilities expected for this system.
The onset of convection occurs when the temperature difference $\Delta T$
across the layer  exceeds a critical value $\Delta T_c$.  For  $\Delta T >
\Delta T_c$,
  the quiescent layer becomes unstable to a periodic pattern of convection
 rolls with wavenumber $ k$.  The stability analysis showed that there is a
well defined region in the $\Delta T - k$ plane, known as the "Busse balloon",
 within which time-independent straight rolls are predicted to be stable.  The
 detailed size and shape of the balloon depends on the Prandtl number
 $\sigma = \nu / \kappa$, where ${\nu}$ is the kinematic viscosity and
${\kappa}$
 the thermal diffusivity.

In this Letter we report experimental results for convection in gaseous
${CO_2}{~~} ({\sigma}{~~} {\simeq} {~~}{0.96})$ in a large aspect
 ratio system ($\Gamma$ = radius / height = 78).  In much of the regime where
 theory predicts time-independent parallel straight rolls, we observed instead
a
 spatially disorganized time-dependent state consisting of many localized
rotating spirals and other defects. There were both right- and left-handed
spirals which rotated clockwise and counterclockwise, respectively. Most
were single armed, but we also observed two- and three- armed spirals and
patches of concentric rolls.  The spirals were {\it{not}} created in pairs,
but rather emerged from and coexisted with surrounding highly disordered
regions in the pattern.  Usually, spirals were created and destroyed in the
 interior of the cell well away from the boundaries.  The overall pattern
appeared to be chaotically time-dependent.   Our results
indicate\refto{BAPSABSTRACT} that this state is representative of
convection in large-${\Gamma }$ systems with ${\sigma \approx 1}$.
This is consistent with early observations on a large-${\Gamma }$ sample
 using liquid Helium,\refto{AB78b} but in that work the patterns were
not visualized.

It is well known that defects and roll curvature give rise to large scale mean
 flows\refto{SZ81,POCHEAU,CROQ2} which can in turn advect the
rolls, leading to complex time dependence.  Such flows have much
 more pronounced effects at low $ \sigma $.  We are unable to visualize
 such flows, but recent numerical simulations\refto{HWXI} suggest
 that they are important for understanding the spiral-defect-chaos state.

Our convection cell consisted of a sapphire top plate and a polished
aluminum bottom plate, each $0.95$ cm thick.  The bottom plate had
 a film heater glued to its lower surface. The lateral boundaries were
constructed of three layers of porous filter paper which was compliant enough
 to allow the cell height to be adjusted by up to 10${\mu}m$ by means of
three piezoelectric stacks. The height $d$ was 568${\mu}{m}$, uniform
to ${\pm}{1}{\mu}{m}$. The cell height and its uniformity were measured
 interferometrically.  The paper sidewalls produced smaller lateral
temperature gradients than the solid ones used previously\refto{BDAC91}
 and caused the pattern to prefer a roll orientation perpendicular to them.
    For most of the results reported here the pressure was $32.7 \pm 0.1$ bar,
regulated to $\pm 0.01$\%. The temperature of the upper surface of the top
 plate was held at $24.00 \pm 0.02 ^{\circ} C$ and regulated to $\pm{2}{mK}$
by means of circulating water maintained at
the same pressure as the gas. The bottom-plate temperature was measured
by means of an embedded thermistor and regulated to ${\pm}0.5{mK}$.
This temperature was varied as the experimental control parameter. This
 protocol
caused the average temperature, and thus the average fluid properties, to
vary with control parameter. We define the reduced temperature difference
 ${\epsilon}{~~} {\equiv}{~~}({\Delta}{T}/{\Delta}{T_{c0}}){-1}$,
where ${\Delta}{T_{c0}}$ is the
critical temperature difference for a fluid having properties corresponding
to those of a sample at the average temperature $\bar T$ .
We found the onset of convection at ${\Delta}{T_c}= 6.622
{\pm} {0.005^{\circ}}{C}$, and used the known temperature
 dependence of the gas properties to obtain ${\Delta}{T_{c0}}({\bar T}) $.
 The characteristic time scale is the vertical thermal diffusion time
${t_v}{~~}{\equiv}{~~}{d^2}/{\kappa}{~~}{\simeq}{~~}{1.3}{s}$.   ${t_v}$
   varied about 15\% over our range of $\bar T$ , while the Prandtl number
 ${\sigma}{~~}{\simeq}{~~}{0.96}$ varied only about 3\% .\refto{NOBCOM}
  The patterns were visualized using the shadowgraph method.\refto{SAC85a}

The states we studied were formed by increasing ${\epsilon}$ from just
below onset to the desired final value in a short
time $(\sim 10{t_v})$. After this quench we waited at least two
horizontal diffusion times
${t_h}{~~}{\equiv}{~~}{\Gamma^2}{t_v}{~~}{\simeq}{~~}{2.2}
{h}$,  for transients to decay. We used this procedure as a matter of
convenience only;  we have obtained similar patterns by increasing ${\epsilon}$
slowly $({t_v} {{d}{\epsilon}\over{dt}} {~~}{\simeq} {~~}10^{-5})$.

Examples of patterns observed for small ${\epsilon}$ are
shown in Fig. 1.   For ${\epsilon}{~~}{{<}\atop{\sim}}{~~}{0.050}$
 we found essentially straight rolls in agreement with theory as shown
 in image 1(a).
The rolls showed a  progressive tendency to become normal to the sidewalls with
increasing
${\epsilon}$, which resulted in strong curvature together with focus
singularities at the sidewalls as shown in image 1(b) for ${\epsilon}
= {0.116}$. This state showed persistent time dependence on timescales of
order ${t_h}$, attributable to the motion of defects, grain boundaries and
foci. This behavior is reminiscent of observations made in a previous
 studies\refto{ACS85, HG87} of water in cylindrical cells.

With increasing ${\epsilon}$, time dependence on shorter time scales
$[{\cal O} (100{t_v})]$ developed in the {\it{interior}} of the cell.
It took the form of transient rotating spiral patches at
${\epsilon}{~~}{\simeq}{~~}{0.4}$, and
for ${\epsilon} {{>}\atop{\sim}} 0.5$ a sea of interacting rotating
spirals and other mobile
defects existed in the interior as shown in image
2(a).\refto{BAPSABSTRACT,CRITPOINT} With further increase in
${\epsilon}$ the area occupied by the foci on the sidewalls
decreased until the cell was filled with what we have termed "spiral defect
chaos",  as shown in image 2(b) for ${\epsilon} = {0.721}$.  Individual
spirals typically rotated several times while translating a distance comparable
to their diameter before being destroyed or suffering a change in the number
of arms.  A common process resulting in a change in the number of
 arms consisted of a dislocation gliding into the spiral core.   Occasionally,
 successive events of this kind left a spiral with opposite handedness to
the original.  Spirals were generally created and destroyed by complicated
 processes involving interactions with the other non-spiral defects in the
pattern, rather than with other spirals.
The timescale for the dynamics decreased with increasing
${\epsilon}$, but a detailed account of the temporal behavior
is beyond the scope of this Letter.

We characterized these patterns using the structure function
${S}({\vec k})$, equal to the time average of the square of the
modulus of the  spatial Fourier transform of the shadowgraph signal.
  ${S}({\vec k})$ provides quantitative information regarding the
spatial scales of the roll patches.
We prefiltered the images by multiplying them by a radial Hanning
function $H(r){~~}{\equiv}{~~}{[{1}+ {cos}({\pi} { r} / {r_0})]/2}$
for ${r \leq {r_0}}$ and ${H(r) }{~~}{\equiv}{~~}0$ for ${r > {r_0}}$.
 We used ${{r_0} = 0.71 {\Gamma}}$ in units of $d$, and averaged 256
measurements of ${S}({\vec k})$ taken at intervals of order severa
l hundred ${t_v}$.   ${S}({\vec k})$ progressed from a few sharp  peaks
to a broad ring as ${\epsilon}$ was increased.
For ${\epsilon} {{>}\atop{\sim}} 0.4$, where many spirals appear,
${S}({\vec k})$ was nearly azimuthally symmetric, i.e., it depended only
 on ${k}{~~}{\equiv}{~~}|{\vec k}|$ and not on ${\vec k}$.   At
each ${\epsilon}$, we performed an azimuthal average in ${\vec k}$-space
to obtain better statistics for  ${S}({k})$. A typical result is presented in
Fig. 3.

In terms of the first two moments
$$
<k>{~~}{\equiv}{~~}{{{\int}|{\vec k}|{S}({\vec k}) {d^2}{\vec k}}\over
{{\int}{S}({\vec k}){d^2}{\vec k}}} =
{{{\int_0^{\infty}}{k^2}{S}({k}){dk}}\over
{{\int_0^{\infty}}{k} {S}({k}) {dk}}}\eqno(1)
$$
and
$$
<k^2>{~~}{\equiv}{~~}{{{\int}|{\vec k}|^2{S} ({\vec k}) {d^2}{\vec k}}\over
{{\int}{S}({\vec k}){d^2}{\vec k}}} =
{{{\int_0^{\infty}}{k^3}{S}({k}){dk}}\over
{{\int_0^{\infty}}{k} {S}({k}) {dk}}}\eqno(2)
$$
of  ${S}({\vec k})$, we define an average wavevector $<k>$ and a
correlation length
$$
{\xi}{~~}{\equiv}{~~}[<k^2> - <k>^2]^{-{1 / 2}}.\eqno(3)
$$
Our results for ${\xi}({\epsilon})$ in units of ${d}$ are presented in Fig. 4.
 The solid line is a fit to a power law of the form ${\xi} = {\xi_0}
{\epsilon^{-\nu}}$, and yields ${\xi_0} = (2.4 \pm {0.1}) {\it d}$ and
 ${\nu} = 0.43 \pm 0.05$. Obviously, ${\xi}$ decreases strongly with increasing
 ${\epsilon}$, and
is only of order a few ${d}$ for ${\epsilon}{{>}\atop{\sim}}{1}$. Under these
conditions a correlation area ${(\pi}{\xi^2})$ occupies less than 0.1\% of the
total cell area.   A statistical description in the infinite-$\Gamma$
limit\refto{CH93} might accurately characterize our experiments for
${\epsilon}{{>}\atop{\sim}}{1}$.  We also show ${\xi}$ for runs at
two other pressures which span a range of fluid parameters.\refto{NOBCOM}
 We find very similar behavior in these cases, with ${\xi}$ tending to
 increase slightly with pressure, for fixed ${\epsilon}$.

The mean wavevector $<k>$ of the pattern decreased with increasing
${\epsilon}$ in such a way that it stayed well within the theoretically
stable region\refto{BC79a, MARCO} (the Busse balloon)  as shown in Fig. 5.
  It has previously been suggested that pattern instabilities,\refto{ACD83}
or the onset of complex time dependence,\refto{PCLeG85, HG87, CROQ2}
occur when the wavevector distortion required to meet the lateral boundary
 conditions forces the pattern to become unstable by {\it locally} exceeding
the stability limits of the uniform infinite pattern.   The analog of this cell
geometry dominated situation in our cell would seem to be the slowly time
dependent state we observed for ${\epsilon}{{<}\atop{\sim}} {0.2}$.
On the other hand, the nucleation and proliferation of spiral defects
 observed at higher ${\epsilon}$
does not fit this picture.   The chaotic state
is already well developed when only a small fraction of the
wavevector distribution lies outside the Busse balloon as shown
in Fig. 3.  It seems more reasonable to explain the broadening of
the wavevector distribution as a consequence of the disordering effects
 of dynamics intrinsic to the pattern, rather than as a response to boundary
 conditions. At higher ${\epsilon}$, the wavevector distribution extends
over both boundaries of the Busse balloon, and we do occasionally
observe, for example, the nucleation of a pair of dislocations in
places where rolls are pushed close together.

The rotation of the spiral defects is a particularly striking feature
of the chaotic state.  Periodic states involving dislocations and spirals,
 both rotating\refto{BDAC91} and nonrotating,\refto{CROQ2}  have
been observed previously in gas convection, but in these cases the spirals
 spanned the experimental cell and were influenced by special lateral
 boundary conditions.  In our experiments, it is clear that the spirals are
coherent structures\refto{CH93}  which emerge as part of the chaotic
dynamics, and are unrelated to the lateral boundaries.  Our results naturally
 raise the question of why the experimental patterns differ so
dramatically  from the theoretical expectations.  It appears
 that the straight-roll state
 is a rather special situation;\refto{CROQ1} the attractor basin of straight
 rolls apparently does not overlap with the initial conditions and boundary
conditions accessible to the experiment.  The spiral defect chaos found at
${\epsilon}{{>}\atop{\sim}} {1}$ is no longer dominated by the
boundaries, and presumably represents a different, unsteady state that
 would persist in the infinite-${\Gamma}$ limit.

We wish to thank Hao-Wen Xi,  J. D. Gunton and J. Vi\~nals for
sharing the results of their simulations, and Yuchou Hu and R. Ecke
 for useful discussions. This research was supported by the
Department of Energy through Grant DE-FG03-87ER13738. S. W. M.
acknowledges support from The Natural Sciences and Engineering
 Research Council of Canada, and E. B. from the Deutsche
 Forschungsgemeinschaft.

\refis{CH93} M. C. Cross and P. C. Hohenberg,  {\it Rev. Mod.
Phys.}, in print.

\refis{CB74} R. M. Clever and F. H. Busse, \jfm 65, 625 1974.

\refis{BDAC91} E. Bodenschatz, J. R. de Bruyn, G. Ahlers and D. S.
Cannell,
 \prl 67, 3078 1991.

\refis{BAPSABSTRACT} S. W. Morris, E. Bodenschatz, D. S.
Cannell and G. Ahlers, \journal Bull. Am. Phys. Soc., 37, 1734 1992.

\refis{HWXI} H. Xi, J. D. Gunton and J. Vi\~nals, preprint.

\refis{CROQ1} V. Croquette, \journal Contemp. Phys., 30, 113 1989.

\refis{CROQ2} V. Croquette, \journal Contemp. Phys., 30, 153 1989.

\refis{AB78b} G. Ahlers and R. P. Behringer, \prl 40, 712 1978.

\refis{SZ81} E. D. Siggia and A. Zippelius, \prl 47, 835 1981.

\refis{NOBCOM} We determined the validity of the Oberbeck-Boussinesq
 (OB) approximation to the Navier-Stokes equations for our fluid by
calculating the non-OB parameter ${\cal P}$ defined by F. H. Busse, \jfm 30,
625 1967.
  ${\cal P}$ varies approximately as ${{\cal P}_c} ( 1+p{\epsilon})$.
  We found ${{\cal P}_c}$ = -0.7, -1.05, -2.1 and $p$ = 1.1, 1.46, 1.79 for
pressures $P$ = 41.5, 32.7 and 25.6 bar, respectively.  It is important to note
that non-OB effects scale as $ 6 {\cal P}^2/R_c$, where $R_c = 1708$,
so that they need not be important when ${\cal P}$ is ${\cal O}(1)$.
See Ref. [\refnd{BDAC91}]. The Prandtl numbers $\sigma$ were
1.06, 0.96 and 0.86 for $P$ = 41.5, 32.7 and 25.6 bar, respectively.
Experimentally, we found that  the chaotic state was qualitatively
the same over this range.

\refis{MARCO} M. A. Dominguez-Lerma, G. Ahlers and
 D. S. Cannell, \journal Phys. Fluids, 27, 856 1984.

\refis{ACS85}  G. Ahlers, D. S. Cannell, and V. Steinberg, \journal
Phys. Rev. Lett., 54, 1373, 1985.

\refis{HG87} M. S. Heutmaker and J. P. Gollub, \pra 35, 242 1987.

\refis{PCLeG85} A. Pocheau, V. Croquette and P. Le Gal, \prl 55, 1094 1985.

\refis{POCHEAU} V. Croquette, P. Le Gal,  A. Pocheau and
R. Guglielmetti,  \journal Europhys. Lett., 1, 393 1986.

\refis{BAPSABSTRACT} We first observed the chaotic state
 under non-Oberbeck-Boussinesq (non-OB) conditions.
See Ref. [\refnd{NOBCOM}] and S. W. Morris, E. Bodenschatz,
D. S. Cannell and G. Ahlers, \journal Bull. Am. Phys. Soc., 37, 1734 1992.

\refis{CRITPOINT} Very recently, M. Assenheimer and V. Steinberg
  (preprint, Feb. 17, 1993) reported spatio-temporally chaotic behavior
 of a many-target pattern, as well as a transition to a many-spiral
 state, near the critical point of $SF_6$.

\refis{BC79a}F.H. Busse and R.M. Clever, {\it J. Fluid Mech.} {\bf 91}, 319
(1979).

 \refis{SAC85a} V. Steinberg, G. Ahlers, and D.S. Cannell, \physs 32, 534 1985.

 \refis{ACD83} G. Ahlers, D.S. Cannell,  and M.A. Dominguez-Lerma, \prl 27,
1225
 1983.

\references

$^*$ Present address: Department of Physics and Erindale College,
University of Toronto, 60 St. George St. Toronto, Ontario, Canada M5S 1A7.

$^\dagger$ Present address: Laboratory of Atomic and Solid
State Physics, Cornell University, Ithaca, New York 14853.

\endreferences

\figurecaptions

\indent Fig. 1.   Examples of patterns observed for
${\epsilon}{{<}\atop{\sim}} {0.2}$. (a) ${\epsilon} = {0.040}$,
nearly perfect straight rolls.  (b) ${\epsilon} = {0.116}$, the global
texture is dominated by curved rolls due to a few focus singularities
on the sidewalls.

\indent Fig. 2.  Pattern sequence as ${\epsilon}$ is increased.  (a)
${\epsilon} = {0.465}$,
 coexistence of sidewall foci with a central chaotic region.
(b) ${\epsilon} = {0.721}$, spiral defect chaos completely fills the cell.

\indent Fig. 3.   The azimuthally and time averaged structure function
of the pattern at
${\epsilon} = {0.465}$.  The dotted lines show the stability boundaries
 for  straight rolls, from the Busse balloon.

\indent Fig. 4.   The correlation length ${\xi}$ vs. ${\epsilon}$. The solid
circles are for a pressure of 32.7 bar (${{\cal P}_c = -1.05,  {\sigma} =
0.96}$,
 see Ref. [\refnd{NOBCOM}]), and the straight line is a fit to
${\xi} = {\xi_0}{\epsilon^{-\nu}}$.  The triangles and squares show
${\xi}$ for 25.6 bar (${\cal P}_c = -2.1,  {\sigma} = 0.86$)  and
41.5 bar (${\cal P}_c = -0.7,  {\sigma} = 1.06$), respectively.

\indent Fig. 5.    A comparison of $<k>$ and width ${\xi}^{-1}$ with
 the stability boundaries (the Busse balloon) predicted for straight
rolls at ${\sigma} = {0.96}$. The dashed curve is the neutral curve.
The solid circles indicate $<k>$, while the horizontal bars extend by  $\pm
{\xi}^{-1}$.

\endfigurecaptions

\endit
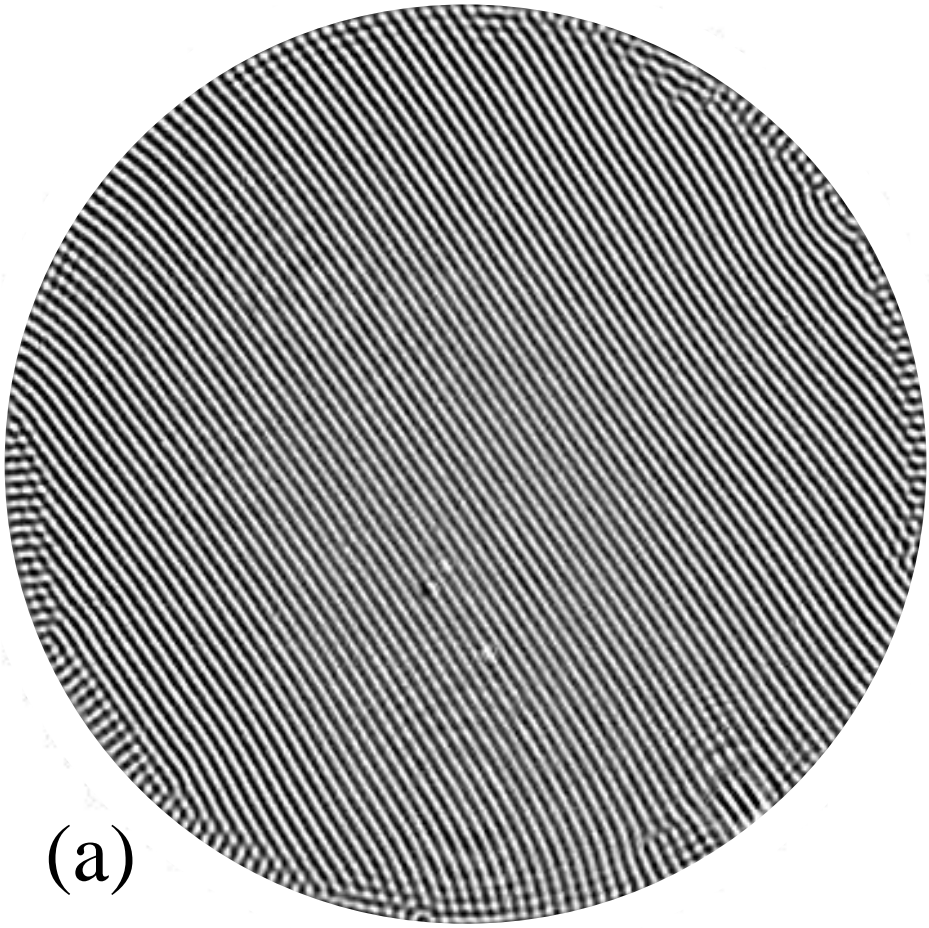
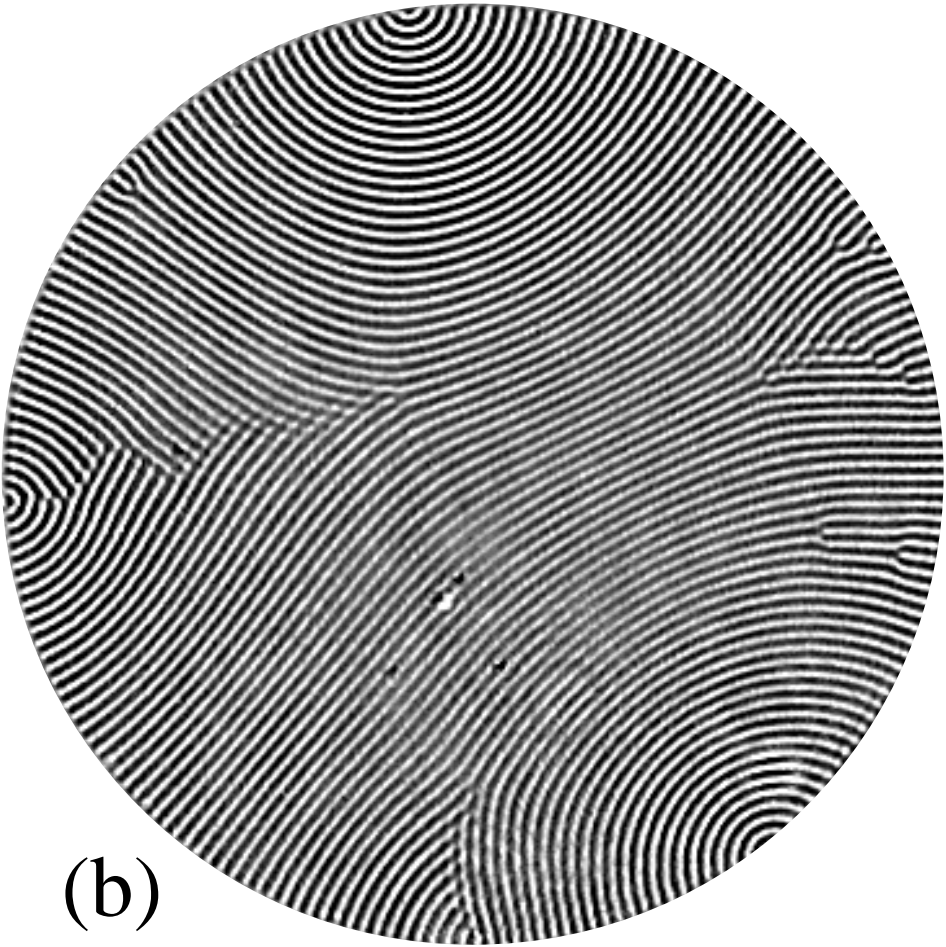
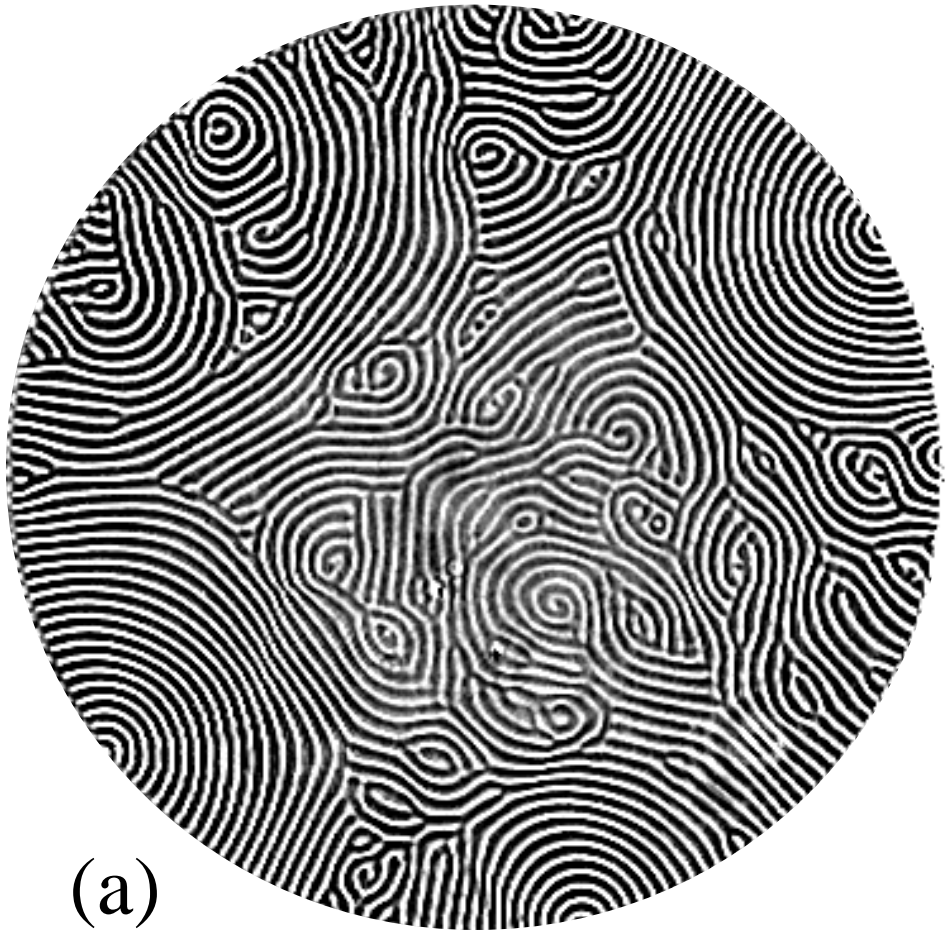
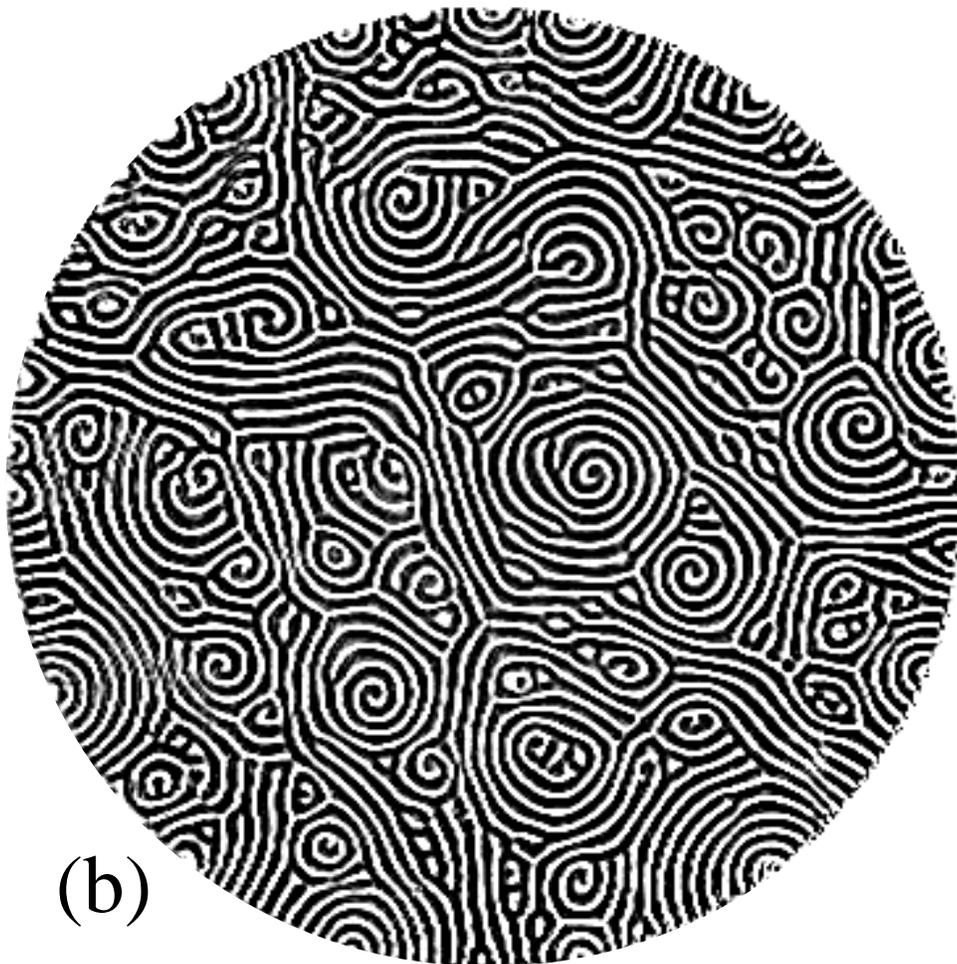
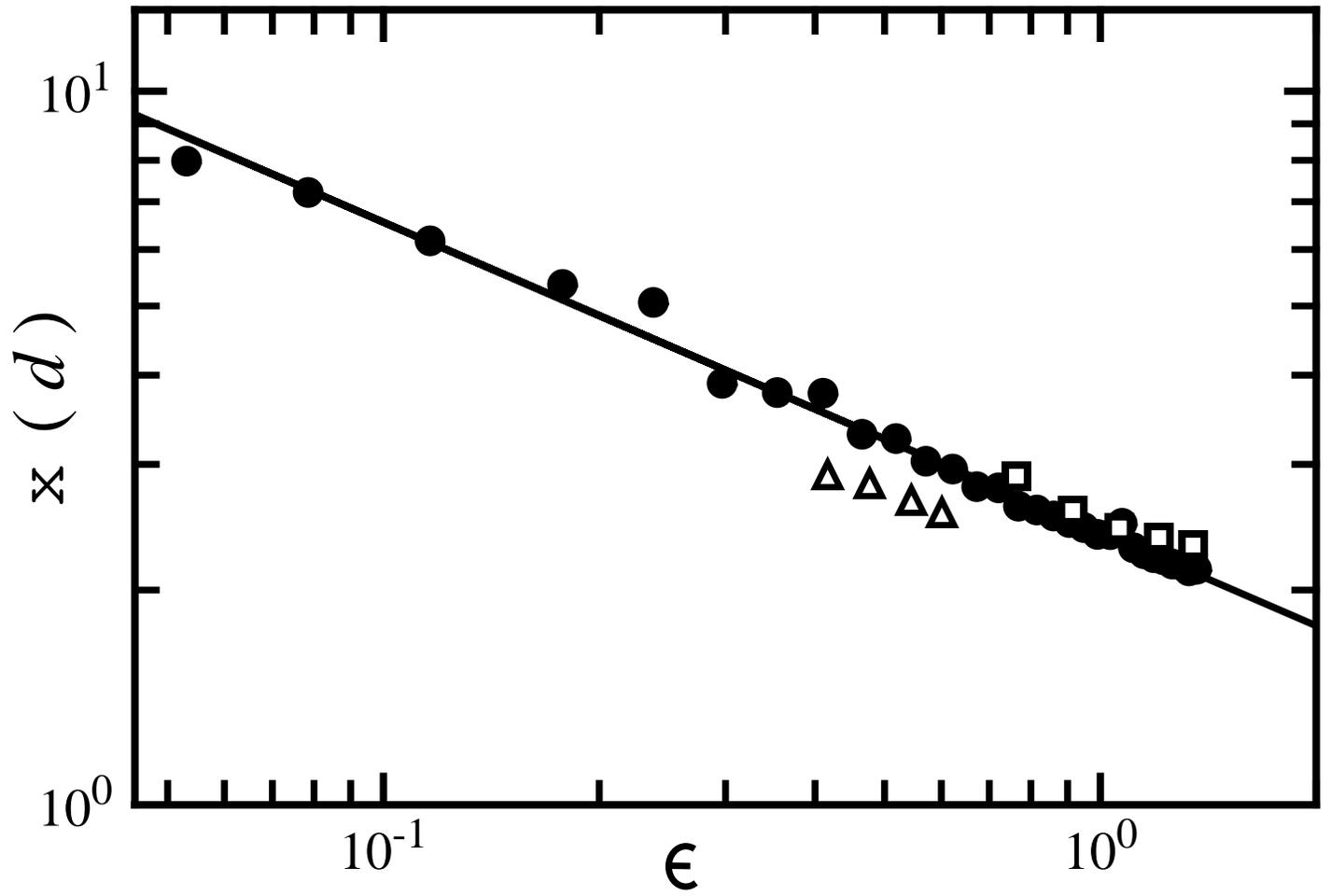
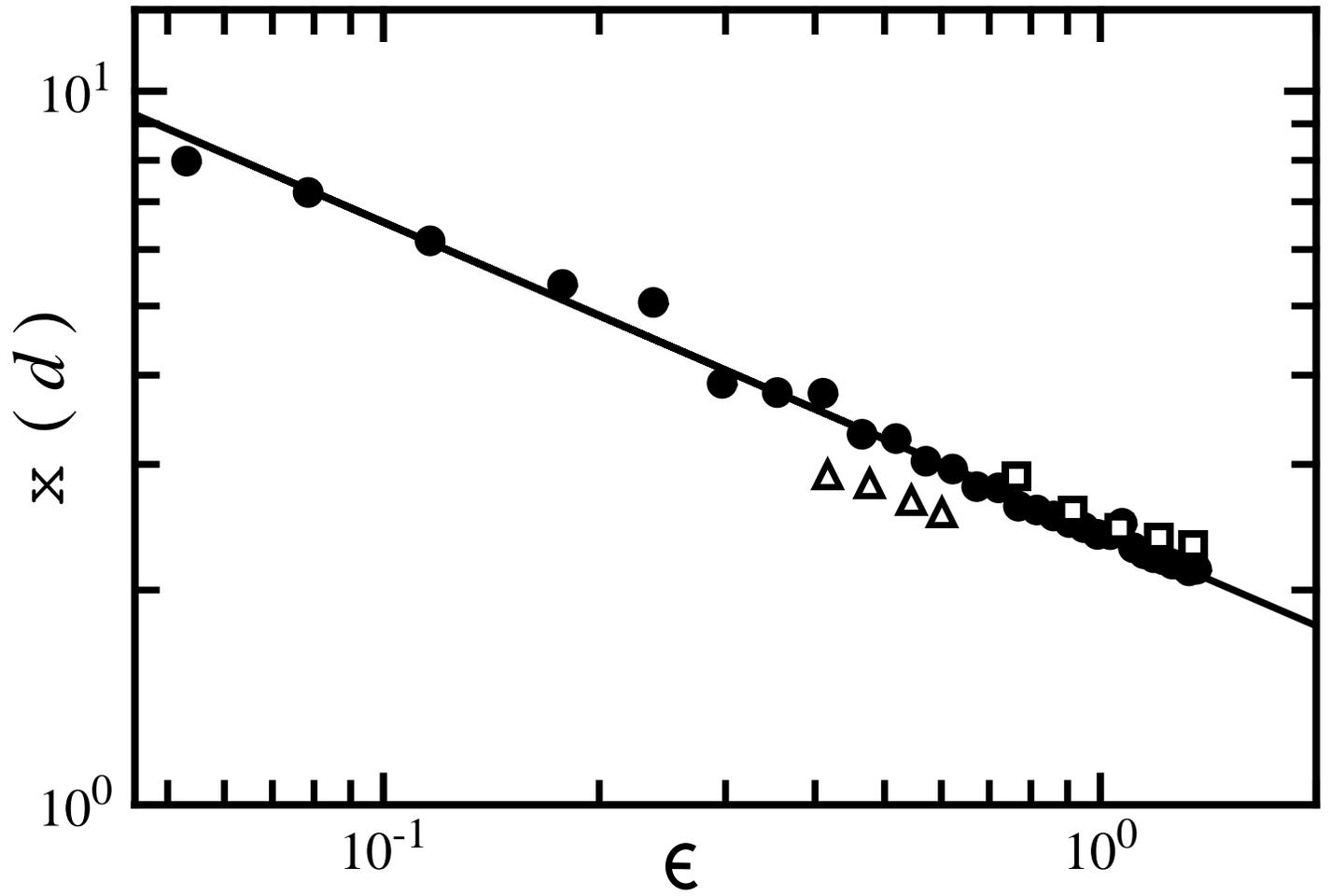
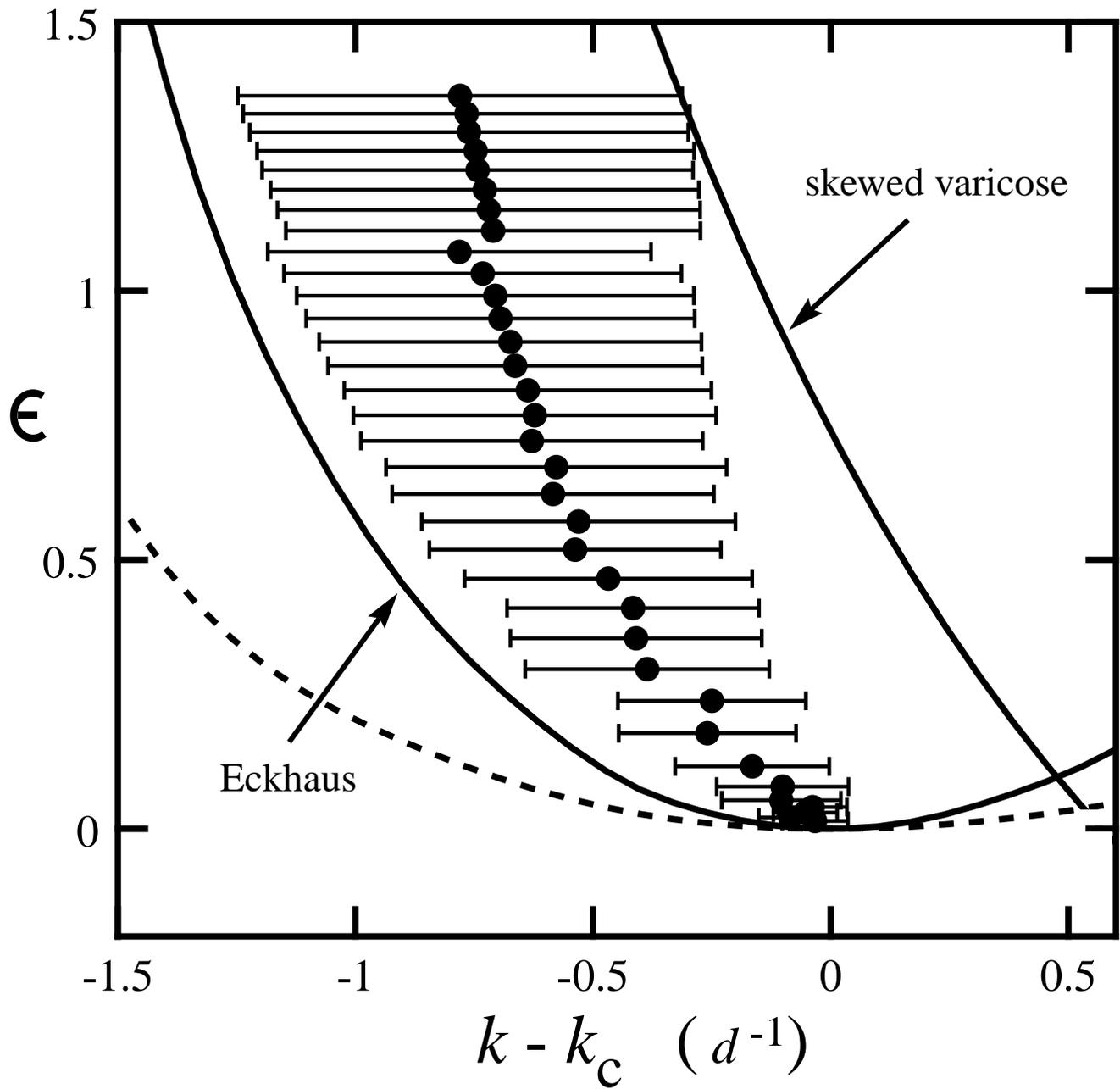
